\documentclass[12pt]{article}
\usepackage{graphicx}

\usepackage{subfig}
\usepackage{color}
\usepackage{epsfig}
\usepackage{feynmf}
\usepackage{url} 
\usepackage{xspace} 

\newcommand\pubdate{\today}

\def\jlab{Thomas Jefferson National Accelerator Facility\\
Newport News,VA 23606, USA}

\def\Title#1{\begin{center} {\Large #1 } \end{center}}
\def\Author#1{\begin{center}{ \sc #1} \end{center}}
\def\Address#1{\begin{center}{ \it #1} \end{center}}

\newcommand\pubblock{\rightline{\begin{tabular}{l}
         \pubdate  \end{tabular}}}

\newenvironment{Presented}{\begin{quotation} \begin{center} 
             PRESENTED AT\end{center}\bigskip 
      \begin{center}\begin{large}}{\end{large}\end{center} \end{quotation}}
\def\Acknowledgements{\bigskip  \bigskip \begin{center} \begin{large}
             \bf ACKNOWLEDGEMENTS \end{large}\end{center}}

\begin{document}

\newcommand{\geo}{$G_{E0}^\Delta$\xspace}
\newcommand{\geii}{$G_{E2}^\Delta$\xspace}
\newcommand{\gmi}{$G_{M1}^\Delta$\xspace}
\newcommand{\gmiii}{$G_{M3}^\Delta$\xspace}
\newcommand{\gpp}{$\gamma^*pp$\xspace}
\newcommand{\gpd}{$\gamma^*p\Delta$\xspace}
\newcommand{\gdd}{$\gamma^*\Delta\Delta$\xspace}
\newcommand{\gpx}{$\gamma^*pX$\xspace}
\newcommand{\gxd}{$\gamma^*X\Delta$\xspace}

\newcommand{\bn}{$B_n$\xspace}
\newcommand{\bnd}{$B_n^\Delta$\xspace}
\newcommand{\DD}{$\Delta(1232)$\xspace}
\newcommand{\qq}{$Q^2$\xspace}
\newcommand{\Al}{$^{27}$Al\xspace}
\newcommand{\C}{$^{12}$C\xspace}

\newcommand{\ff}{form-factor\xspace}
\newcommand{\ffs}{form-factors\xspace}
\newcommand{\tff}{transition form-factor\xspace}
\newcommand{\tffs}{transition form-factors\xspace}

\newcommand{\etal}{{\it et al.}\xspace}

\begin{titlepage}
\pubblock

\vfill
\Title{Accessing the Elastic Form-Factors of the $\Delta(1232)$ Using the Beam-Normal Asymmetry}
\vfill
\Author{ Mark Macrae Dalton}
\Address{\jlab}
\vfill
\begin{abstract}
The beam-normal single-spin asymmetry, $B_n$, exists in the scattering of high energy electrons, polarized transverse to their direction of motion, from nuclear targets. To first order, this asymmetry is caused by the interference of the one-photon exchange amplitude with the imaginary part of the two-photon exchange amplitude. Measurements of $B_n$, for the production of a \DD resonance from a proton target, will soon become available from the Qweak experiment at Jefferson Lab and the A4 experiment at Mainz. The imaginary part of two-photon exchange allows only intermediate states that are on-shell, including the $\Delta$ itself. Therefore such data is sensitive to \gdd, the elastic form-factors of the $\Delta$. This article will introduce the form-factors of the $\Delta$, discuss what might be learned about the elastic form-factors from these new data, describe ongoing efforts in calculation and measurement, and outline the possibility of future measurements.
\end{abstract}
\vfill
\begin{Presented}
Twelfth Conference on the Intersections of Particle and Nuclear Physics\\
Vail, Colorado, USA, May 19-24, 2015
\end{Presented}
\vfill
\end{titlepage}
\def\thefootnote{\fnsymbol{footnote}}
\setcounter{footnote}{0}

\section{Introduction}

The nucleon is a complex object and has a rich excitation spectrum.  Much has been learned about properties of the nucleon, such as polarizabilities, magnetic moments, form factors and partonic structure.  
There is much information on the lowest lying nucleon resonances, including their masses, quantum numbers, widths, major decay modes and branching fractions.  
However, new and different data is required to truly test understanding of internal degrees of freedom of the excited states.

Elastic \ffs of hadrons are interesting observables since they contain information on internal properties, such as the distribution of charge, that can be compared with calculations.
There is a significant program underway, using electron beams, to measure the elastic  \ffs of the proton, the neutron and even the pion, which continues to ever higher momentum-transfer and precision. 
These three particles $(p,n,\pi)$ are currently all that can be measured elastically, so \ff studies are extended to measurements in the transition from one state to another.  These transition \ffs conflate information from both the initial and final states, which makes extracting properties of one of the individual states more difficult. 

A particularly interesting and well studied object is the \DD resonance.
The \DD is the lowest lying excitation of the nucleon system and is experimentally well separated from other resonant and non-resonant scattering processes.  On the other hand, it is very short lived, so studying its elastic properties is extremely challenging.  There is a relationship between the elastic \ffs of the \DD and an observable called the beam-normal single-spin asymmetry which may allow the experimental study of these elastic \ffs.
This paper will describe this relationship and assess the status of current experimental and theoretical efforts to obtain the elastic \ffs.

\section{$\Delta$ elastic form-factors}

Being a spin-3/2 object, the \DD resonance has 4 \ffs which represent the structure beyond that of a pure point-like spin-3/2 particle.  These can be written as $G_{E0}(Q^2)$, $G_{M1}(Q^2)$, $G_{E2}(Q^2)$, and $G_{M3}(Q^2)$ which correspond to the charge, magnetic, quadrupole and octupole \ffs respectively. 

The magnetic moment and quadrupole moment in the transition $p\rightarrow\Delta(1232)$ have been measured very precisely in \DD photo-production. However, very few of the \DD elastic properties have been explicitly measured.  One notable exception is the elegant determination of the magnetic moment of the \DD through a $\gamma$-transition within the resonance~\cite{Kotulla:2002cg}.  Unfortunately, this extraction is dominated by a theoretical uncertainty and is consistent with 0.   The elastic properties may be studied through lattice calculations of QCD and information on the transverse quark charge densities and non-spherical deformations have been found~\cite{Alexandrou:2009hs}.  For a comprehensive review of electromagnetic excitation of the \DD, see Pascalutsa \etal\cite{Pascalutsa:2006up}.

\section{Beam-normal single-spin asymmetry $B_n$}

A beam-normal single-spin asymmetry, often abbreviated as transverse asymmetry, (henceforth \bn in this article) is a scattering asymmetry observed when a beam of electrons, polarized transverse to the propagation direction, is scattered in the plane normal to the polarization direction.  The asymmetry is time-reversal odd, and time-reversal invariance forces it to vanish for the Born term, where a single photon is exchanged.  The leading order contribution is then from an interference between the one-photon exchange amplitude and the imaginary part of the two-photon exchange amplitude.
\begin{equation}
B_n=\frac{\sigma_\uparrow-\sigma_\downarrow}{\sigma_\uparrow+\sigma_\downarrow}
=\frac{2\textrm{Im}(T_{2\gamma}\cdot T_{1\gamma}^*)}{|T_{1\gamma}|^2}
\end{equation}

The asymmetry is small, at the level of a few to a hundred parts-per-million (ppm), since it suppressed by $m_e/E_e$ to polarize the ultra-relativistic electron perpendicular to its direction of motion, and by $\alpha_\textrm{em}$ in the exchange of the second photon; 
\begin{equation}
B_n\sim \alpha_{\textrm{em}}\frac{m_e}{E_e}\sim10^{-6} - 10^{-5}.
\end{equation}

\subsection{$B_n$ in elastic electron-proton $ep$ scattering}

\bn has been measured in elastic electron scattering from a proton in a number of experiments at various kinematics~\cite{Androic:2011rh,Armstrong:2007vm,Maas:2004pd,Wells:2000rx} including an extremely precise new result, at very forward angle, by the Qweak Experiment~\cite{Waidyawansa:2013yva}.  Calculations of these quantities were performed integrating over the doubly virtual Compton scattering tensor modeled using $\gamma^* N \rightarrow \pi N$ amplitudes~\cite{Pasquini:2004pv}.
These calculations do a reasonable job of describing the data qualitatively over a wide phase space but there is some quantitative disagreement, particularly with the Qweak data.  These disagreements with the data may arise from intermediate states that are not included using the $\pi N$ amplitudes.

In the case of very forward angles, calculations have been done using the optical theorem and parameterizations for the measured total photo-production cross sections on the proton~\cite{Afanasev:2004pu,Gorchtein:2006mq} which describe the Qweak data very well.

\subsection{$B_n$ in elastic electron-nucleus scattering}

\bn in elastic electron-nucleus scattering has recently been measured on $^4$He, $^{12}$C and $^{208}$Pb~\cite{Abrahamyan:2012cg} by the HAPPEX and PREX collaborations, at very forward angle.  The forward angle, optical-theorem calculations have been extended to nuclei and a very simple dependence on  $A/Z$ is predicted~\cite{Gorchtein:2008dy}.  The calculations describe the data very well, except for the  observation of an anomaly in scattering off $^{208}$Pb, where the asymmetry is extremely small, up to a 30 $\sigma$ difference from naive expectations.

\begin{figure}[htb!]
\centering
\includegraphics[width=0.7\textwidth]{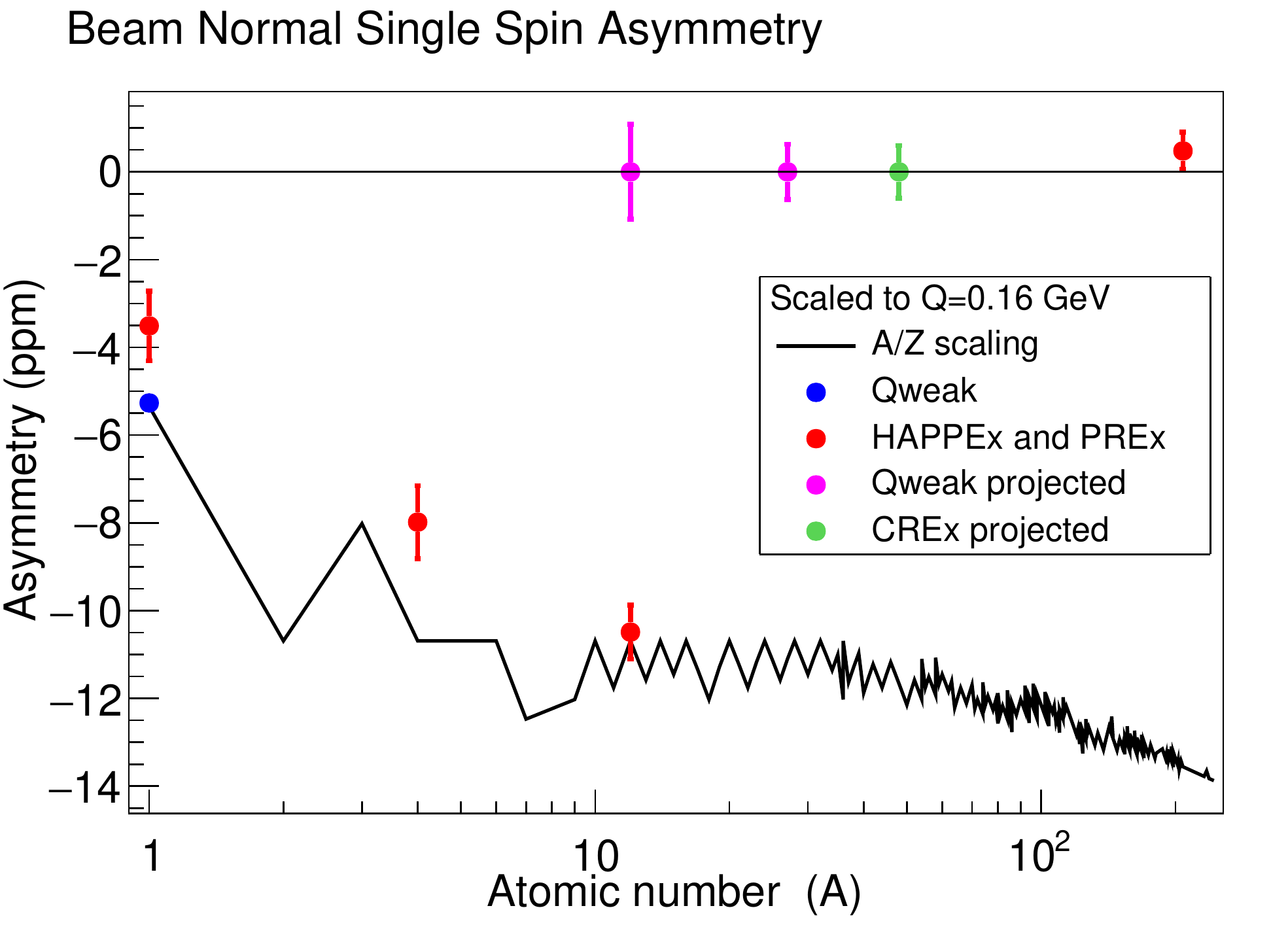}
\caption{\bn versus atomic number for existing results from the HAPPEX, PREX and Qweak Experiments along with projected statistical uncertainties for results under analysis by Qweak and a potential measurement by the CREX Experiment.  The HAPPEX and PREX data (red)~\cite{Abrahamyan:2012cg} have been scaled linearly in $Q$ to a value of $Q=0.16$~GeV to match the Qweak data (blue)~\cite{Waidyawansa:2013yva}.  The black solid line is a scaling of the Qweak hydrogen result by $A/Z$, following Ref.~\cite{Gorchtein:2008dy}.}
\label{fig:nuclear_bn}
\end{figure}

Figure \ref{fig:nuclear_bn} shows \bn versus atomic number for existing and projected future measurements.  In order to show all the data on the same plot, the HAPPEX and PREX data have been scaled linearly in $Q$ to a value of $Q=0.16$~GeV, matching the Qweak data.  A note of warning; the linear scaling in $Q$ is a phenomenological observation from calculations and is likely more appropriate for changes in angle than changes in energy.  The curve shows a scaling of the Qweak hydrogen result by $A/Z$ to all stable nuclei.  The figure shows that $QA/Z$ scaling is quite well satisfied by the data except for the very heaviest nucleus, which is likely due to Coulomb distortion effects.

Qweak has taken \Al and \C data, also at $Q=0.16$~GeV, which are currently under analysis---with publication expected in 2016.  These new data will allow precision testing of the $A/Z$ scaling prediction up to $A=27$.  The Calcium Radius Experiment (CREX)~\cite{prop_CREX} should produce \bn data on $^{48}$Ca and perhaps even $^{40}$Ca data within the next decade.

In summary, calculations for \bn in elastic processes are basically under good control, aside from the obvious disagreement for $^{208}$Pb.  At larger angles where the optical theorem cannot be used, calculations using only $\pi N$ amplitudes seem sufficient to reproduce the dominant features.

\subsection{$B_n$ in $ep\rightarrow e\Delta(1232)$}
\label{sec:eped}

The transverse asymmetry in \DD production, henceforth abbreviated \bnd, is more difficult to calculate since data for particular vertices do not exist.  In fact, this is the very foundation of the current interest in this process---one may try to use the overall asymmetry to infer properties of the individual vertices. 

\begin{figure}[htb]
\centering
\includegraphics[width=0.99\textwidth]{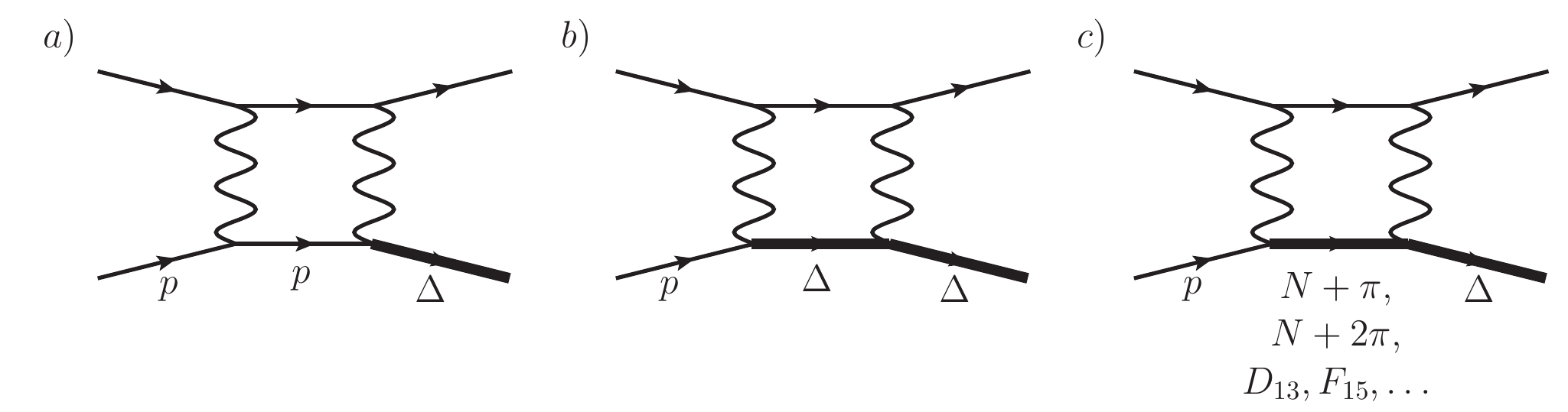}
\caption{Feynman diagrams for \DD electro-production through 2-photon exchange.  This is the lowest order process that can produce a  non-zero \bn.  The hadronic part in the loop may be a) the proton, b) the \DD, or c) another excited state of the proton.}
\label{fig:feynman}
\end{figure}

Consider the Feynman diagrams in Figure~\ref{fig:feynman}, which show a lepton exciting a proton into a \DD resonance through the exchange of two photons.  The hadronic part in the loop may be a) the proton, b) the \DD, or c) another excited state of the proton, $X$.  It can be seen that in order to calculate the amplitude for diagram a), data for the vertices \gpp and \gpd are required, while for diagram b) data for the vertices \gpd and \gdd are required, and diagram c) requires data from \gpx and \gxd, where $X$ represents all other potential proton excitations.  

From these diagrams it can be seen than \bnd, through the 2-$\gamma$ production of \DD, is sensitive to \gdd \ffs at first order.  The challenge lies in using measurements of \bnd to extract information on \gdd without losing too much sensitivity due to the extraction method and theoretical unknowns.  In accounting for the total \bnd asymmetry in \DD production, calculations need to be done for the 5 different vertices mentioned above.

The \gpp vertex is just the proton elastic \ff.  This has been well measured to high \qq and does not present a problem.  The \gpd and \gpx vertices are the familiar \tffs, $p-$\DD and $p-X$ respectively, which have also been measured quite well for the \DD and extracted for the higher lying states using partial wave analysis of large data sets.  The \gdd vertex is the \DD elastic \ff.  This is the quantity of interest and will be discussed in more detail in the next section.

The \gxd vertex, is the biggest unknown in a calculation of \bnd.  At the moment there are not even order-of-magnitude estimates for the size of this contribution.  Such calculations appear to be possible using the constituent quark model and theoretical efforts are underway to pursue this. 
This will be discussed in a later section.

\subsubsection{\DD intermediate state}

Any measurement of \bnd integrates over all intermediate states and all intermediate kinematics, in particular of the 2 exchanged photons.  Thus it is not possible to extract \ff information at any particular \qq.  The most promising way to extract information on \gdd from \bnd is to assume particular \ffs and then to compute \bnd.  By varying the input \gdd, while remaining consistent with the measured \bnd, it is possible to constrain the shape and magnitude of the \ffs.   A promising initial parametrization is from fits to lattice calculations~\cite{Alexandrou:2009hs}, of the form, 
\begin{eqnarray}
G_{E0}(Q^2) & = & \frac{1}{(1+Q^2/\Lambda^2_{E0})^2} \\
G_{M1}(Q^2) & = & G_{M1}(0)e^{-Q^2/\Lambda^2_{M1}} \\
G_{E2}(Q^2) & = & G_{E2}(0)e^{-Q^2/\Lambda^2_{E2}} \\
G_{M3}(Q^2) & = & 0,
\end{eqnarray}
with parameters that are approximately, $\Lambda^2_{E_0}\approx1.2$, $G_{M_1}(0)\approx2.4$, $\Lambda^2_{M_1}\approx1$, $G_{E_2}(0)\approx-1$, and $\Lambda^2_{E_2}\approx1$.  The electric \ff is parametrized as a dipole, while the magnetic and quadrupole \ffs are allowed to fall faster, as an exponential.  Any contribution from the octupole \ff is neglected since it is consistent with zero in current lattice calculations.  One could imagine using a different parametrization that makes the sensitivity to the static properties more apparent, for example the charge radius,
\begin{equation}
G_{E_0}(Q^2)  =  1 - \frac{1}{6}R^2_\Delta Q^2 + \dots \\
\end{equation}

\begin{figure}%
    \centering
    \subfloat{{\includegraphics[width=0.45\textwidth]{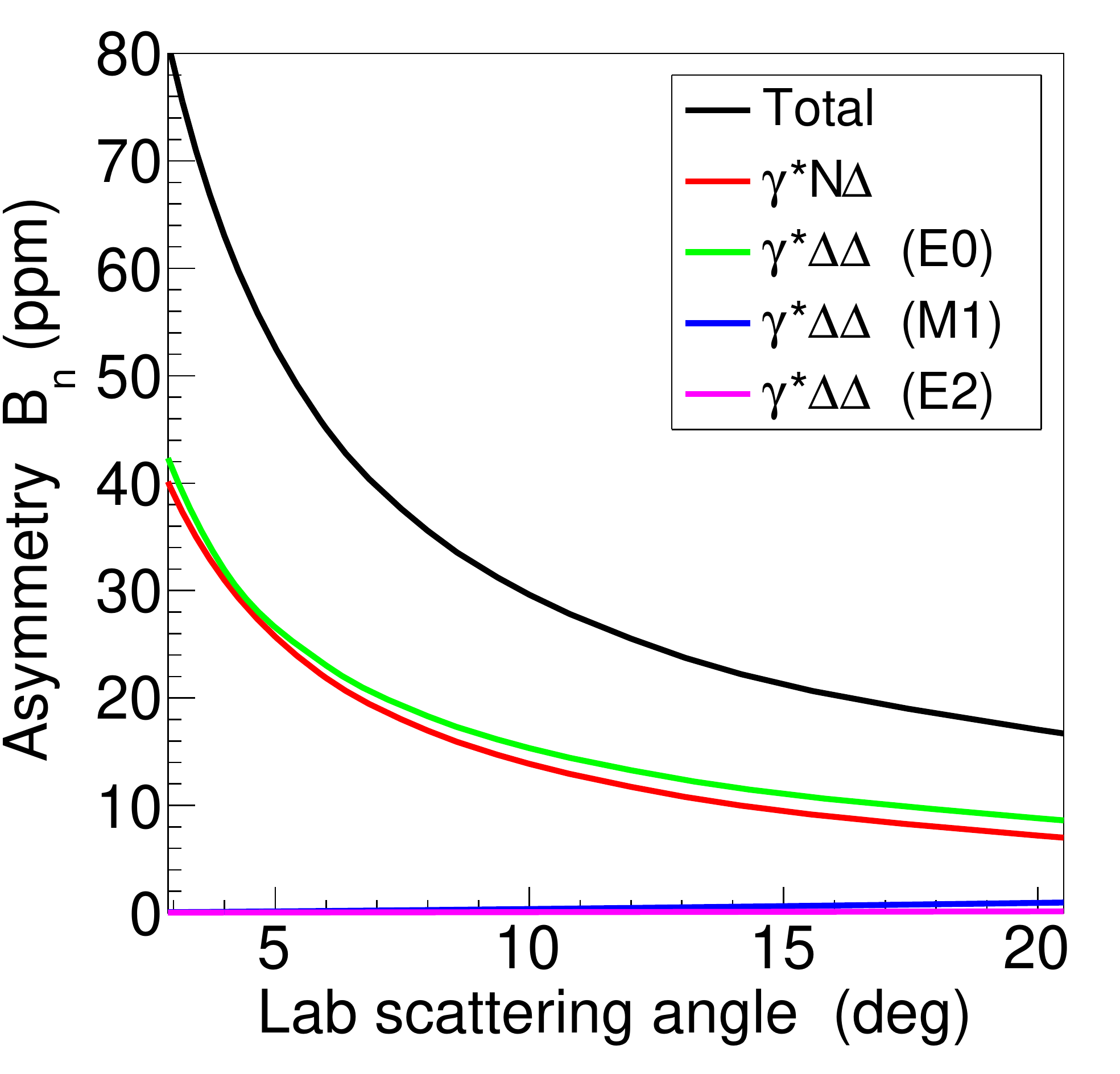} }}%
    \qquad
    \subfloat{{\includegraphics[width=0.45\textwidth]{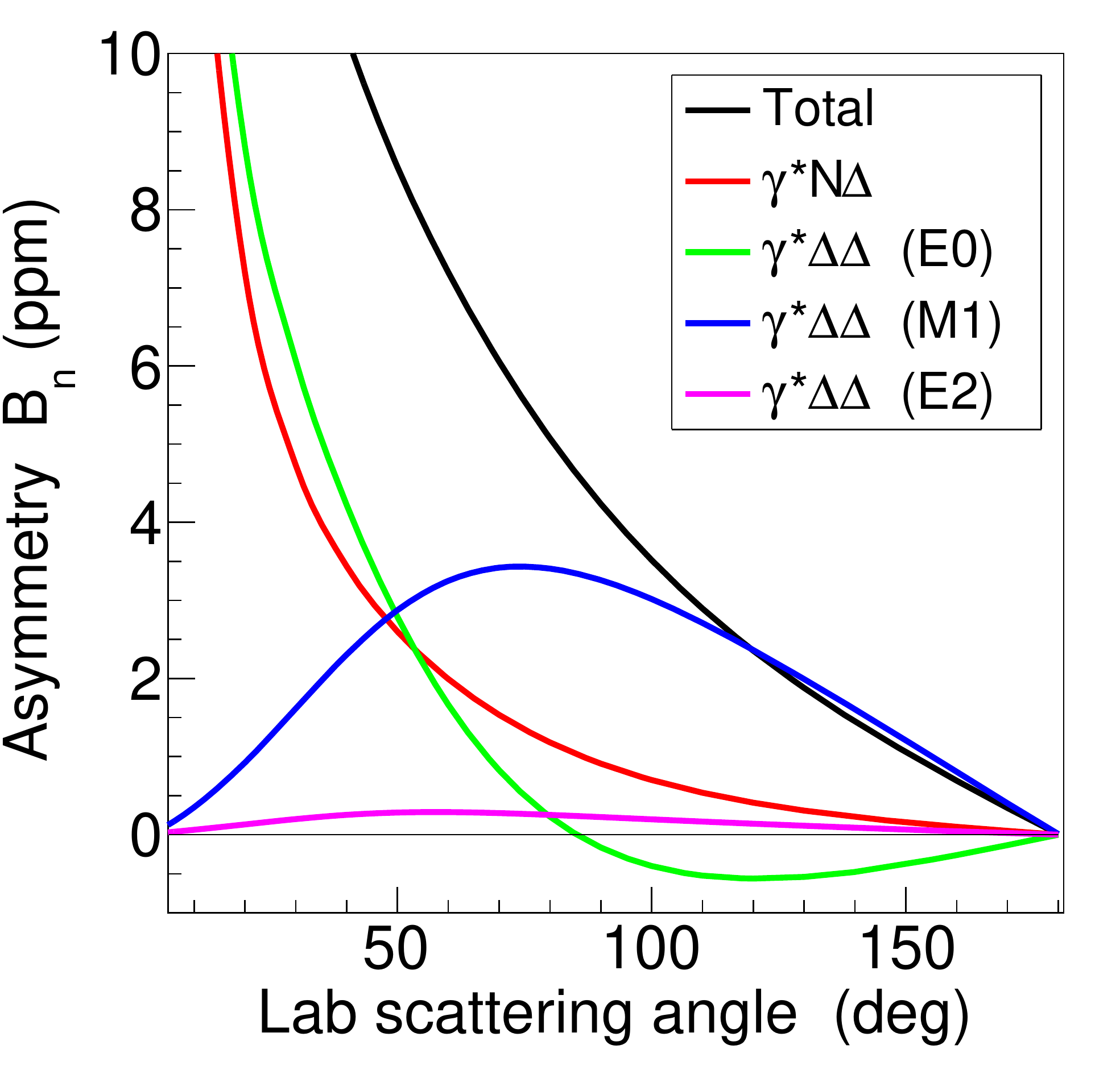} }}%
    \caption{Calculations of $B_n$ in $ep\rightarrow e\Delta(1232)$ at one of the Qweak beam energies, E=1.16 GeV, by B. Pasquini~\cite{pasquini_private}.  The left panel shows the large asymmetries at small angles, while the right panel  emphasizes the smaller asymmetries over the full angular range.  The total contribution is also separated into contributions from the elastic intermediate state and the \DD intermediate state for the 3 largest \ffs.}%
    \label{fig:pasquini}
\end{figure}

Figure \ref{fig:pasquini} shows a calculation of \bnd in \DD production at a beam energy of 1.16~GeV by Barbara Pasquini \etal~\cite{pasquini_private}.  Only the elastic and \DD intermediate states have been included.
The parametrization described above is used for \gdd.  
At very forward angles, the asymmetry is large and has equal contributions from the elastic and \DD intermediate states, where the \DD component is dominated by the \geo \ff.  At backward angles, the \gmi \ff from the \DD intermediate state dominates, but the asymmetry is much smaller.  The \geii \ff looks to be too small to access using this method.  This calculation along with a more comprehensive set of sensitivity studies is expected to be published in the near future.

\subsubsection{Higher mass intermediate states}

The higher mass diagrams (Fig.~\ref{fig:feynman} c) contain 2 vertices, \gpx and \gxd, where $X$ is any intermediate state except the nucleon or \DD.  The \gpx vertex, the proton transition \ffs, have been extensively measured and the amplitudes extracted. 
The \gxd vertex on the other hand is unknown with no current prospects of being extracted from data.  This particular issue is currently the biggest barrier to interpretation of \bnd measurements in terms of \gdd amplitudes.  Without taking them into account, contributions from higher mass intermediate states will be ascribed to the \gdd amplitudes.   Naively, it seems like significant contributions are possible.  Consider the case of \bn for elastic scattering off the proton~\cite{Pasquini:2004pv}, where the elastic intermediate state makes essentially negligible contribution and almost all of the strength comes from higher mass intermediate states.  Theoretical studies in this regard are now beginning, with the initial goal of estimating the order of magnitude of the contribution.  

\begin{figure}[htb]
\centering
\includegraphics[width=0.55\textwidth]{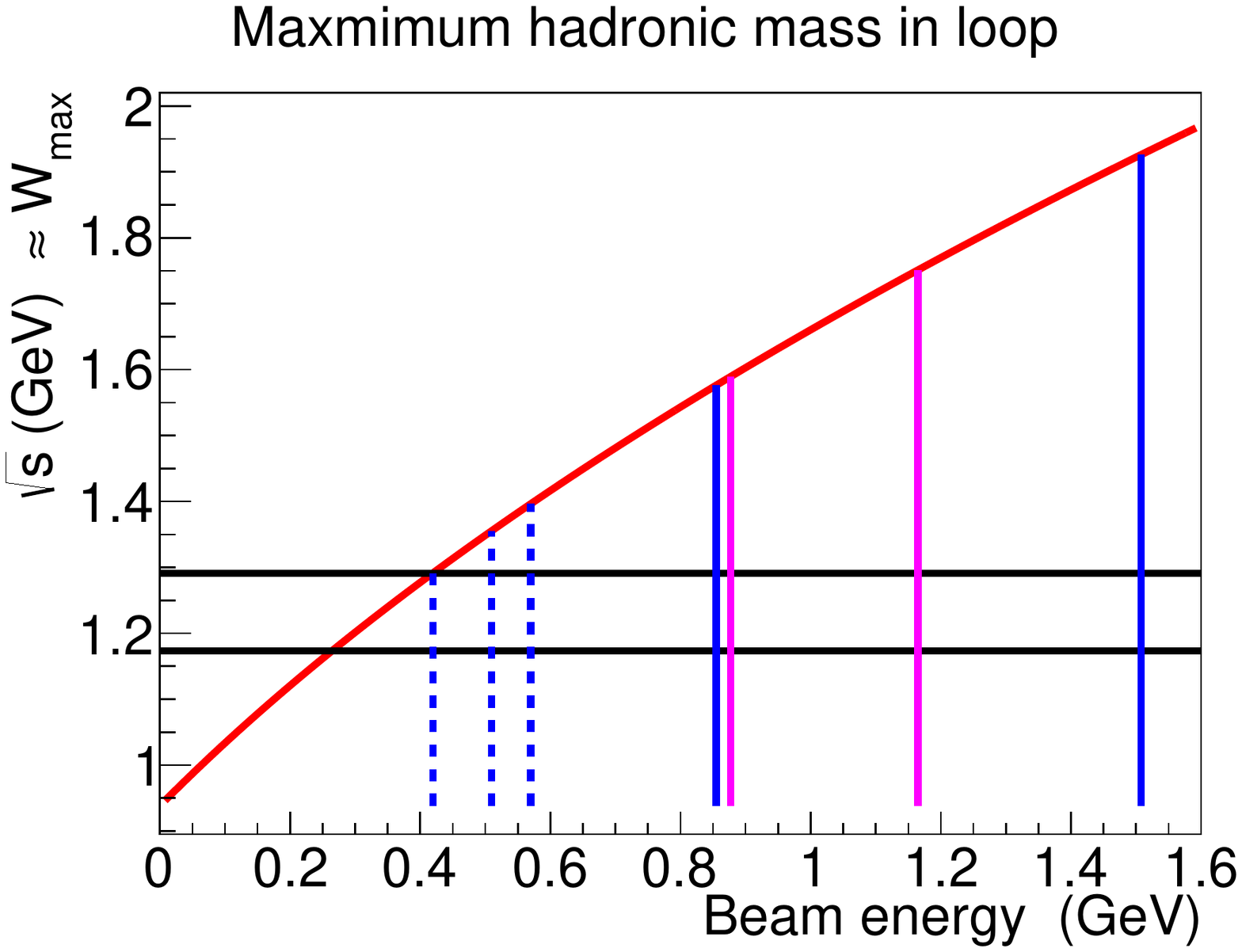}
\caption{The maximum hadronic mass in the loop versus the energy of the electron beam.  Vertical lines indicate the data under analysis by Qweak (magenta) and Mainz (blue), described in the next section.  Dashed lines indicate data for which an extraction may not be possible. Horizontal lines give the 1-$\sigma$ width of the \DD.}
\label{fig:beam}
\end{figure}

Experimentally, the most promising method of controlling higher mass intermediate states is to use a low beam energy.  Figure~\ref{fig:beam} shows  $\sqrt{s}$, which is approximately the maximum mass that can appear within the loops of Fig.~\ref{fig:feynman}, versus beam energy.  By keeping a low beam energy, and by extracting \bnd as a function of energy, contributions from higher mass states can be studied and minimized.

\section{Experimental measurements of \bnd}

There is not much data available on \bnd right now, but data taken as part of studies during 2 recent parity-violation experiments, and currently under analysis, should be available during 2016.

\subsection{Qweak at Jefferson Lab}

The Qweak experiment measured \bn in the \DD region at two beam energies, 855 MeV and 1160 MeV.  The very thick target and forward kinematics lead to a large background contribution from elastic scattering that radiates into the \DD kinematics.  The uncertainty in the extraction of \bnd from these data is completely dominated by the knowledge of relative fractions elastic and \DD production.   A preliminary result for the 1160 MeV was presented at CIPANP15~\cite{Nuruzzaman:2015vba}.  This result is significantly non-zero and agrees with calculation presented above, although it has a relatively large uncertainty which stems almost exclusively from uncertainty in the elastic background contribution.  An improved analysis is underway using a more sophisticated treatment of radiative effects, following Dasu~\etal~\cite{Tsai:1971qi_misc,Dasu:1993vk}. 

\subsection{A4 at Mainz}

The PVA4 experiment at Mainz took data on \bn in $ep$ scattering, intending to look at the elastic component.  Reference~\cite{Maas:2004ta} shows a nice example of the calorimeter spectrum showing that, at an energy of 855 MeV, the \DD is sufficiently separated to make an extraction of \bnd viable.  Such data has been taken at $\theta = 30^\circ-40^\circ$  with beam energies of 855 MeV and 1508 MeV.  The statistical uncertainties for these data in the \DD region are 2.1 ppm and 5.1 ppm respectively, which is promising considering that \bnd is predicted to be $\sim30$ ppm for the lower beam energy~\cite{pasquini_talk}.

Beam normal asymmetry data were also taken at 420 MeV, 510 MeV, and 570 MeV at $\theta = 30^\circ-40^\circ$  and 420 MeV at $\theta = 140^\circ-150^\circ$.  Extracting \bnd from these data might not be possible, since at  low scattered electron energies the calorimetry technique has difficulty detecting and resolving inelastic processes.  If it can be extracted, these data would be very interesting since the low beam energies, and the range of beam energies at the same angle, should help to constrain contributions from higher order intermediate states.  The solitary backward angle measurement would be sensitive to \gmi.

\section{Future work}

\subsection{Theoretical input}

These studies require significant theoretical input if we are to establish a robust method for measuring the \DD  \ffs.  My thoughts on what is required, in my own rough order of priority:

\begin{itemize}
  \item Study the impact of higher mass intermediate states on \bnd.  To truly extract \gdd  from \bnd would require such calculations to be under control.  For the time being, it would be interesting just to get an estimate of the size of such effects.  
  \item Study the sensitivity of \bnd to the parameters of \gdd.  We would ultimately like to obtain data of sufficient precision to test models of the \DD  \ffs.  For example how quickly do they fall with \qq and what is the charge radius?   Sensitivities to the parameters would be required to determine how precise new experimental data must be.
  \item Extend the existing calculations of \bnd to include energies from \DD threshold to $\sim$1 GeV.  Such a set of calculations would be needed to plan future measurements of \bnd and for doing  radiative corrections.
\end{itemize} 

\subsection{Potential future measurements}

Should theoretical studies determine that information on the \gdd  \ffs can be reliably and unambiguously extracted from \bnd data, then a program of dedicated future measurements would be quite promising.  The A1 facility at Mainz might be able to execute such a program.  There are plans to do parity-violation experiments on heavy nuclei in order to measure their neutron radii, which would necessarily require the development of all the appropriate experimental techniques.  Expanding to include \DD production should be relatively easy.  Another example is the proposed SOLID Experiment~\cite{prop_SOLID} at Jefferson Lab.  However, SOLID would be limited to $\theta = 20^\circ-35^\circ$ and, due to scheduling of multiple experiments simultaneously at Jefferson Lab, might not be able to go to the low beam energies appropriate for this measurement.  Other examples of potential facilities are the Cornell-BNL FFAG-ERL Test Accelerator and the Low Energy Recirculating Facility (LERF) at Jefferson Lab.

\section{Conclusion}

The beam-normal single-spin asymmetry in \DD production is sensitive to the \DD elastic \ffs, \gdd.
Measurement of beam asymmetries is a robust, mature technique and, in a dedicated experiment, the tens of ppm sized asymmetries typical of \bnd could be measured with low systematic error in a relatively short amount of time.
Angle dependence should allow the separation of the \ffs; forward angles probe \geo while backward angles probe \gmi.
Lower beam energies are most desirable for \bnd measurements since they minimize effects from higher mass intermediate states.
A future dedicated program of such measurements could be possible with reasonable beam time at various facilities worldwide, particularly where parity violation experiments will have already developed all the necessary infrastructure, such as A4 at Mainz and SOLID at Jefferson Lab.
Theoretical studies are needed to determine sensitivity of the asymmetry, \bnd, to the assumed form and size of \gdd.  This would help to determine what the ideal kinematics and precision of any future measurements would be.

\Acknowledgements

Many thanks to Barbara Pasquini and Carl Carlson for helpful discussions, existing and future calculation efforts.  Thanks also to Sebastian Baunack for information on the existing A4 Mainz data.  Acknowledgements to the Qweak Collaboration, of which I am member.

 \bibliographystyle{nar}
\bibliography{dalton_bibtex_delta,dalton_bibtex_transverse,dalton_bibtex_parity,dalton_bibtex_proposals,dalton_bibtex_general}

\begin{thebibliography}{10}

\bibitem{Kotulla:2002cg}
Kotulla, M. et al. (2002)
{\em Phys. Rev. Lett.} {\bf 89}, 272001.

\bibitem{Alexandrou:2009hs}
Alexandrou, C., Korzec, T., Koutsou, G., Lorce, C., Negele, J.~W., et al.
  (2009)
{\em Nucl.Phys.} {\bf A825}, 115--144.

\bibitem{Pascalutsa:2006up}
Pascalutsa, V., Vanderhaeghen, M., and Yang, S.~N. (2007)
{\em Phys.Rept.} {\bf 437}, 125--232.

\bibitem{Androic:2011rh}
Androic, D. et al. (2011)
{\em Phys.Rev.Lett.} {\bf 107}, 022501.

\bibitem{Armstrong:2007vm}
Armstrong, D.~S. et al. (2007)
{\em Phys.Rev.Lett.} {\bf 99}, 092301.

\bibitem{Maas:2004pd}
Maas, F., Aulenbacher, K., Baunack, S., Capozza, L., Diefenbach, J., et al.
  (2005)
{\em Phys.Rev.Lett.} {\bf 94}, 082001.

\bibitem{Wells:2000rx}
Wells, S. et al. (2001)
{\em Phys.Rev.} {\bf C63}, 064001.

\bibitem{Waidyawansa:2013yva}
Waidyawansa, B.~P. (2013)
{\em AIP Conf.Proc.} {\bf 1560}, 583--587.

\bibitem{Pasquini:2004pv}
Pasquini, B. and Vanderhaeghen, M. (2004)
{\em Phys.Rev.} {\bf C70}, 045206.

\bibitem{Afanasev:2004pu}
Afanasev, A.~V. and Merenkov, N. (2004)
{\em Phys.Lett.} {\bf B599}, 48.

\bibitem{Gorchtein:2006mq}
Gorchtein, M. (2007)
{\em Phys.Lett.} {\bf B644}, 322--330.

\bibitem{Abrahamyan:2012cg}
Abrahamyan, S. et al. (2012)
{\em Phys.Rev.Lett.} {\bf 109}, 192501.

\bibitem{Gorchtein:2008dy}
Gorchtein, M. and Horowitz, C.~J. (2008)
{\em Phys.Rev.} {\bf C77}, 044606.

\bibitem{prop_CREX}
Mammei, J. et al.
\url{http://www.jlab.org/exp_prog/proposals/12/PR12-12-004.pdf} (2012).

\bibitem{pasquini_private}
Pasquini, B.
Private communication.

\bibitem{Nuruzzaman:2015vba}
Nuruzzaman (2015)
{\em arXiv:1510.00449}.

\bibitem{Tsai:1971qi_misc}
Tsai, Y.-S.
SLAC-PUB-0848 (1971).

\bibitem{Dasu:1993vk}
Dasu, S. et al. (1994)
{\em Phys. Rev.} {\bf D49}, 5641--5670.

\bibitem{Maas:2004ta}
Maas, F. et al. (2004)
{\em Phys.Rev.Lett.} {\bf 93}, 022002.

\bibitem{pasquini_talk}
Pasquini, B.
\url{http://conference.kph.uni-mainz.de/mamiandbeyond/site/program/} (2009)
MAMI and Beyond, 30 March-3 April 2009, Budenheim (Mainz).

\bibitem{prop_SOLID}
Souder, P. et al.
\url{http://www.jlab.org/exp\_prog/proposals/10/PR12-10-007.pdf} (2009).

\end{thebibliography}

\end{document}